\documentclass[twoside,twocolumn,9pt]{article}
\usepackage{extsizes}
\usepackage[super,sort&compress,comma]{natbib}
\usepackage[version=3]{mhchem}
\usepackage[left=1.5cm, right=1.5cm, top=1.785cm, bottom=2.0cm]{geometry}
\usepackage{balance}
\usepackage{widetext}
\usepackage{times,mathptmx}
\usepackage{sectsty}
\usepackage{graphicx}
\usepackage{lastpage}
\usepackage[format=plain,justification=raggedright,singlelinecheck=false,font={stretch=1.125,small,sf},labelfont=bf,labelsep=space]{caption}
\usepackage{float}
\usepackage{fancyhdr}
\usepackage{fnpos}
\usepackage[english]{babel}
\usepackage{array}
\usepackage{droidsans}
\usepackage{charter}
\usepackage[T1]{fontenc}
\usepackage[usenames,dvipsnames]{xcolor}
\usepackage{setspace}
\usepackage[compact]{titlesec}
\usepackage{srctex}
\usepackage{hyperref}

\definecolor{cream}{RGB}{222,217,201}

\begin{document}

\pagestyle{fancy}
\thispagestyle{plain}
\fancypagestyle{plain}{

\fancyhead[C]{\includegraphics[width=18.5cm]{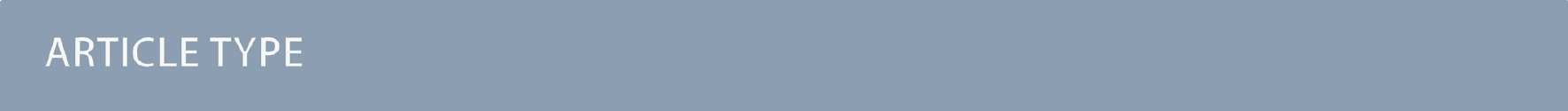}}
\fancyhead[L]{\hspace{0cm}\vspace{1.5cm}\includegraphics[height=30pt]{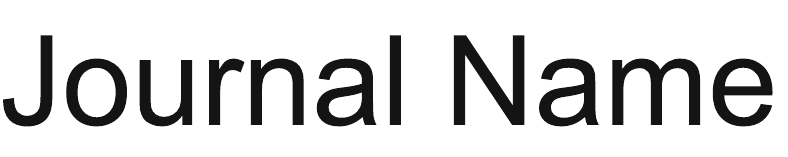}}
\fancyhead[R]{\hspace{0cm}\vspace{1.7cm}\includegraphics[height=55pt]{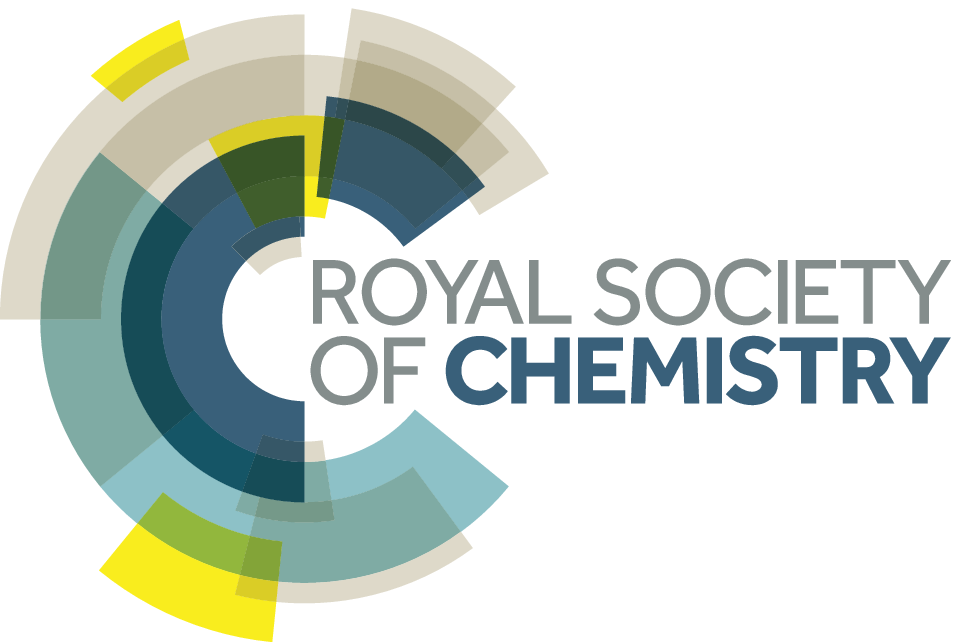}}
\renewcommand{\headrulewidth}{0pt}
}

\makeFNbottom
\makeatletter
\renewcommand\LARGE{\@setfontsize\LARGE{15pt}{17}}
\renewcommand\Large{\@setfontsize\Large{12pt}{14}}
\renewcommand\large{\@setfontsize\large{10pt}{12}}
\renewcommand\footnotesize{\@setfontsize\footnotesize{7pt}{10}}
\makeatother

\renewcommand{\thefootnote}{\fnsymbol{footnote}}
\renewcommand\footnoterule{\vspace*{1pt}%
\color{cream}\hrule width 3.5in height 0.4pt \color{black}\vspace*{5pt}}
\setcounter{secnumdepth}{5}

\makeatletter
\renewcommand\@biblabel[1]{#1}
\renewcommand\@makefntext[1]%
{\noindent\makebox[0pt][r]{\@thefnmark\,}#1}
\makeatother
\renewcommand{\figurename}{\small{Fig.}~}
\sectionfont{\sffamily\Large}
\subsectionfont{\normalsize}
\subsubsectionfont{\bf}
\setstretch{1.125} 
\setlength{\skip\footins}{0.8cm}
\setlength{\footnotesep}{0.25cm}
\setlength{\jot}{10pt}
\titlespacing*{\section}{0pt}{4pt}{4pt}
\titlespacing*{\subsection}{0pt}{15pt}{1pt}

\fancyfoot{}
\fancyfoot[LO,RE]{\vspace{-7.1pt}\includegraphics[height=9pt]{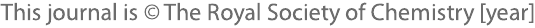}}
\fancyfoot[CO]{\vspace{-7.1pt}\hspace{13.2cm}\includegraphics{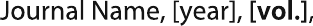}}
\fancyfoot[CE]{\vspace{-7.2pt}\hspace{-14.2cm}\includegraphics{RF}}
\fancyfoot[RO]{\footnotesize{\sffamily{1--\pageref{LastPage} ~\textbar  \hspace{2pt}\thepage}}}
\fancyfoot[LE]{\footnotesize{\sffamily{\thepage~\textbar\hspace{3.45cm} 1--\pageref{LastPage}}}}
\fancyhead{}
\renewcommand{\headrulewidth}{0pt}
\renewcommand{\footrulewidth}{0pt}
\setlength{\arrayrulewidth}{1pt}
\setlength{\columnsep}{6.5mm}
\setlength\bibsep{1pt}

\makeatletter
\newlength{\figrulesep}
\setlength{\figrulesep}{0.5\textfloatsep}

\newcommand{\topfigrule}{\vspace*{-1pt}%
\noindent{\color{cream}\rule[-\figrulesep]{\columnwidth}{1.5pt}} }

\newcommand{\botfigrule}{\vspace*{-2pt}%
\noindent{\color{cream}\rule[\figrulesep]{\columnwidth}{1.5pt}} }

\newcommand{\dblfigrule}{\vspace*{-1pt}%
\noindent{\color{cream}\rule[-\figrulesep]{\textwidth}{1.5pt}} }

\makeatother

\twocolumn[
  \begin{@twocolumnfalse}
\vspace{3cm}
\sffamily
\begin{tabular}{m{4.5cm} p{13.5cm} }

\includegraphics{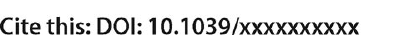} & \noindent\LARGE{\textbf{Self-assembly of the decagonal quasicrystalline order in simple three-dimensional systems}} \\
\vspace{0.3cm} & \vspace{0.3cm} \\

 & \noindent\large{Roman Ryltsev,$^{\ast}$\textit{$^{ab}$} Boris Klumov,\textit{$^{bc}$} and Nikolay Chtchelkatchev\textit{$^{bd}$}} \\

\includegraphics{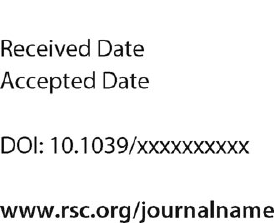} & \noindent\normalsize{Using molecular dynamics simulations we show that a one-component system can be driven to a three-dimensional decagonal (10-fold) quasicrystalline state just by purely repulsive, isotropic and monotonic interaction pair potential with two characteristic length scales; no attraction is needed. We found that self-assembly of a decagonal quasicrystal from a fluid can be predicted by two dimensionless effective parameters describing the fluid structure. We demonstrate stability of the results under changes of the potential by obtaining the decagonal order for three particle systems with different interaction potentials, both purely repulsive and attractive, but with the same values of the effective parameters. Our results suggest that soft matter quasicrystals with decagonal symmetry can be experimentally observed for the same systems demonstrating the dodecagonal order for an appropriate tuning of the effective parameters.} \\

\end{tabular}

 \end{@twocolumnfalse} \vspace{0.6cm}

  ]

\renewcommand*\rmdefault{bch}\normalfont\upshape
\rmfamily
\section*{}
\vspace{-1cm}


\footnotetext{\textit{$^{a}$~Institute of Metallurgy, UB RAS, 620016, Amundsena 101, Ekaterinburg, Russia. E-mail: rrylcev@mail.ru}}
\footnotetext{\textit{$^{b}$~L.D. Landau Institute for Theoretical Physics, RAS, 142432, Ac. Semenov 1-A, Chernogolovka, Russia.}}
\footnotetext{\textit{$^{c}$~High Temperature Institute, RAS, 125412, Izhorskaya 13/2, Moscow, Russia}}
\footnotetext{\textit{$^{d}$~Moscow Institute of Physics and Technology,141700, Institutskiy per.9, Dolgoprudny, Moscow Region, Russia}}



\section{Introduction}

Since their discovery in the 1980s,~\cite{Shechtman1984PRL} quasicrystals (QCs) have demanded increasing attention due to their remarkable physical properties. QCs have been observed, both experimentally and computationally, not only in metallic alloys,~\cite{Tsai2008SciTechAdvMat} but also in molecular systems~\cite{Wasio2014Nature,Dubois2011PhilMag,Johnston2010JChemPhys} and soft matter~\cite{Zeng2004Nature,Fischer2011PNAS,Hayashida2007PRL,Talapin2009Nature,Zaidouny2014SoftMatt,Ungar2005SoftMatt} that suggests the microscopic mechanism of QC formation is common for these systems.


 \begin{figure*}
  \centering
  \includegraphics[width=0.9\textwidth]{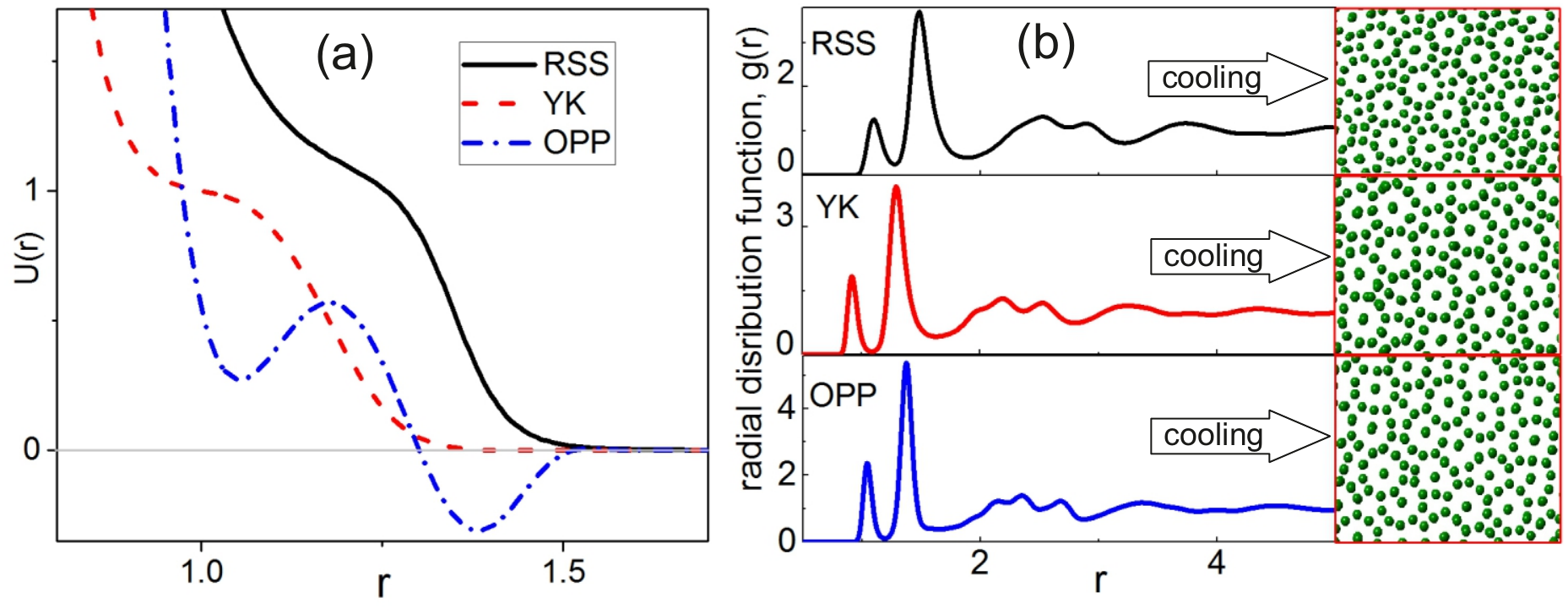}\\
 \caption{Different potentials with similar effective parameters exhibit similar structure. a) Three types of two-length-scale pair potentials that we investigated: RSS -- Repulsive Shoulder System potential \cite{Fomin2008JCP}, YK -- Yoshida-Kamakura potential \cite{Yoshida1976ProgTheorPhys}, OPP -- modified Oscillating Pair Potential from \cite{Mihalkovich2012PRB,Englel2015Nature}. The dimensionless units are used for $U$ and $r$ (see Sec.~\ref{sec:methods});  b) Fluid state radial distribution functions of systems with potentials in (a) and typical snapshots of their low-temperature solid phases with decagonal symmetry. Parameters: RSS -- $\sigma=1.37$, $\rho=0.474$, $T=0.11$; YK -- $\rho=0.74$, $T=0.09$; OPP - $\rho=0.62$, $T=0.16$. Effective parameters $(\lambda$, $\phi$) are (1.35, 0.11) for RSS; (1.4, 0.106) for YK; (1.33, 0.15) for OPP.}
  \label{fig1:RDF}
\end{figure*}

 It has been shown by molecular dynamics simulations that the QC order does not require an extraordinary interaction but it can be obtained for simple one-component systems with isotropic pair potentials.~\cite{Smith1991PRB,Dzugutov1993PRL,Jagla1998PRE,Skibinsky1999PRE,Quandt1999PRB,Roth2000PRE,Keys2007PRL,Engel2007PRL,Archer2013PRL,Barkan2011PRB,Dotera2014Nature,Engel2014PRL,Englel2015Nature} However, most of these investigations have been restricted to quite special case of two-dimensional systems.

 Stability of three-dimensional (3D) one-component QCs has been predicted using density functional theory in different systems whose effective isotropic pair interactions involve more than one length scale, e.g., colloid-polymer mixtures~\cite{Denton1998PRL} and simple-metal-like systems.~\cite{Denton1997EPL,Denton1997PRB} But there are only two examples of computational self-assembly of 3D QCs, dodecagonal (12-fold) and icosahedral, through the simulations of systems with such multi-scale potentials.~\cite{Dzugutov1993PRL,Englel2015Nature} These potentials were designed to mimic oscillating effective interactions in metals; they have attractive wells specially tuned to encourage formation of QC order. We show that attraction is not necessary and the only existence of two interparticle length-scales is the sufficient condition for QC formation. Doing molecular dynamics simulations of simple one-component system with purely repulsive soft-matter-like potential we observe solid phases with 3D decagonal (10-fold) QC order at certain parameters domain. We find that self-assembly of decagonally ordered solid from a fluid phase can be predicted by two dimensionless structural parameters of the fluid.  The parameters reflect the existence of two effective interparticle distances (bond lengthes) originated from two-length-scale nature of interaction potential. These are the ratio between effective bond lengthes, $\lambda$, and the fraction of short-bonded particles $\phi$. We have estimated the optimal values of these parameters for decagonal order and validated them on three systems with essentially different two-length scales potentials. Adjusting the systems parameters to obtain appropriate $\lambda$, $\phi$ values for the fluid phases we observe self-assembly of decagonal QC solids for all the systems considered. That suggests the proposed criterion is robust under change of potential and may be applicable to any system with two-length-scale interaction.

 Two-length-scale potentials like we study are widely used~\cite{Ryzhov2003PRE,Xu2006JPhysCondMatt,Xu2009JChemPhys,Xu2011JChemPhys,Vilaseca2011JNonCrystSol,Kumar2005PRE,Ryltsev2013PRL,Gribova2009PRE,Oliveira2008JChemPhys,Buldyrev2009JPhysCondMatt,Buldyrev2009JPhysCondMatt}; they qualitatively describe effective interactions in molecular systems,~\cite{Mishima1998Nature,Yan2008PRE} soft matter systems~\cite{Likos2001PhysRep,Watzlawek1999PRL,Likos2002JPhysCondMatt,Prestipino2009SoftMatt,Rechtsman2006PRE} and metallic alloys~\cite{Lee1981Book,Mitra1978JPhysC,Mihalkovich2012PRB,Dubinin2009CentEuropJPhys,Dubinin2014JNonCrystSol} where QC formation has been experimentally observed but not completely accounted for.

\section{Methods\label{sec:methods}}

\subsection{Interatomic potentials}

We investigate by the molecular dynamics simulations one-component 3D systems of particles interacting via three different two-length-scale potentials (see Fig~\ref{fig1:RDF}a). The main one is the repulsive shoulder system (RSS) potential:~\cite{Fomin2008JCP}
\begin{equation}
U_{\rm rss}(r)=\varepsilon\left({d}/{r}\right)^{n}+\varepsilon {\rm n_f}\left[2k_0\left(r-\sigma \right)\right],
\end{equation}
where ${\rm n_f}(x)=1/[1+\exp(x)]$, $\varepsilon$ -- is the unit of energy, $d$ and $\sigma$ are ``hard''-core and ``soft''-core diameters. We take $n=14$, $k_0=10$, and $\sigma\in(1.3,1.45)$. These parameters produce phase diagrams with polymorphous transitions,~\cite{Fomin2008JCP} water-like anomalies~\cite{Gribova2009PRE} and glassy dynamics.~\cite{Ryltsev2013PRL} Hereafter we use dimensionless quantities: $\tilde{{\bf r}}\equiv {\bf r}/d$, $\tilde U=U/\varepsilon$, temperature $\tilde T=T/\epsilon$, density $\tilde{\rho}\equiv N d^{3}/V$, and time $\tilde t=t/[d\sqrt{m/\varepsilon}]$, where $m$ and $V$ is the molecular mass and system volume correspondingly. As we will only use these reduced variables, we omit the tildes.

 To validate our main conclusions we also use purely repulsive Yoshida-Kamakura potential
 \begin{equation}
  U_{\rm yk}(r)=\exp(a(1-r)-b(1-r)^m\ln(r))
 \end{equation}
  with $a=0.5$, $b=120$ and oscillating pair potential (OPP):
\begin{equation}
U_{\rm opp}(r)=1/r^{15}+a\exp(-(r/b)^m)\cos(kr-\varphi)
\end{equation}
with $a=0.5$, $b=1.45$, $m=20$, $k=14.4$, $\varphi=17.125$. The former was proposed to analyze high pressure crystal structures~\cite{Yoshida1976ProgTheorPhys} and the latter is slightly modified potential which was fist introduced in \cite{Mihalkovich2012PRB} and then used to simulate icosahedral QCs.~\cite{Englel2015Nature} In contrast to OPP from,~\cite{Englel2015Nature}  we have just replaced pre-cosine power factor by exponential one to suppress oscillations after second minimum (to restrict the system by only two characteristic lengh-scales).

\subsection{Simulation details}

For MD simulations, we have used $\rm{LAMMPS}$ molecular dynamics package.~\cite{lammps} The system of $N=5000-20000$ particles was simulated under periodic boundary conditions in Nose-Hoover NVT and NPT ensembles. This amount of particles is enough to obtain satisfactory diffraction patterns to study (quasi)crystal symmetry (see Fig.~\ref{fig2:snapshots}). Taking larger system requires too much calculation time necessary to QC equilibration. The MD time step was $\delta t=0.003-0.01$ depending on system temperature. It is nearly the maximum possible time step value that provides energy conservation at chosen thermodynamic conditions.

 The RSS was studied in wide density region of $\rho\in(0.35-0.75)$. For YK and OPP systems, which were mostly used to validate generality of the results, the density regions were chosen to obtain appropriate values of dimensionless parameters (see below). They were $\rho\in(0.67-0.74)$ for YK and $\rho\in(0.6-0.62)$ for OPP.

To study solid phases we cooled the system starting from a fluid in a stepwise manner and completely equilibrated at each step. The time dependencies of temperature, pressure and configurational energy were analyzed to control equilibration.

\subsection{Preparation and analysis of solid phases}

Investigation of solid phases has been performed in several stages. At first ``express'' stage we localize system parameters corresponding to different solid state structures. For this, we cool small system ($N\sim 5000$) starting from a fluid state down to transition to a solid and analyse radial distribution functions $g(r)$ (see Fig.~\ref{fig1:RDF}b and inserts in Fig.~\ref{fig2:snapshots}(i,ii)) and bond order parameters $q_l$~\cite{Steinhardt1981PRL,Steinhardt1983PRB,KlumovPU10,KlumovPRB11,Hirata2013Science} (see Fig.~\ref{fig4:q4q6}).  At the second stage, for ``interesting'' parameters, we study the larger system of $N\sim20000$  annealing it for a long time (up to $3\cdot10^8$ MD steps) to reduce (quasi)crystal defects and strains.~\cite{Dzugutov1993PRL, Englel2015Nature} Then the final structure is studied by analyzing snapshots and diffraction patterns.

It is important to note that we determine solid phases in the absence of any data about phase diagrams of the systems under investigation (except some partial data for the RSS \cite{Fomin2008JCP} ). Calculation of such diagrams requires comparison of thermodynamic potentials for (infinity) number of all possible phases to detect the most stable one. As a rule, researchers restrict themselves by considering of some finite set of possible solid structures.~\cite{Fomin2008JCP,Prestipino2009SoftMatt} But even relatively simple system (like we study) can demonstrate complex solid structures (QC, for example) which are hardly predicted in advance and so this approach may fail. An instructive example is the RSS model for which an attempt to calculate phase diagram has been performed.~\cite{Fomin2008JCP} In that case, for $\sigma=1.35$, $\rho\in (0.5, 0.55)$ no simple stable phases were detected. It is exactly the same density range where we find decagonal QC phase. In this connection, the method of "experimental" determination of solid structures by their self-assembly from fluid phase is more useful for our purposes. The solid phases obtained by this method may not be true ground state structures and they may be stable at only certain temperature range. We see that observed phases are stable during the annealing and this is enough for our purposes. The detailed analysis of phase stability and phase diagram determination is the matter of separate work.

Note that the presence of the periodic boundary conditions impedes the relaxation of QC structure and growth of relatively large phason-free QC. This problem is avoidable at simulation of systems with attraction potential because a solid phase can be condensed from the gas one (see, for example, Ref.~\cite{Englel2015Nature}). In the case of pure repulsion one has to obtain a solid phase from  a liquid one with using periodic boundary conditions. In that case the preparation of an satisfactory QC structures is the challenging task. For example, the relaxation of structures represented in Fig.~\ref{fig2:snapshots} required about  $3\cdot 10^8$ MD steps that took a month of calculations on 64 processors in parallel for each configuration.

\subsection{Local orientational order analysis}

\begin{figure}[t]
  \includegraphics[width=0.99\columnwidth]{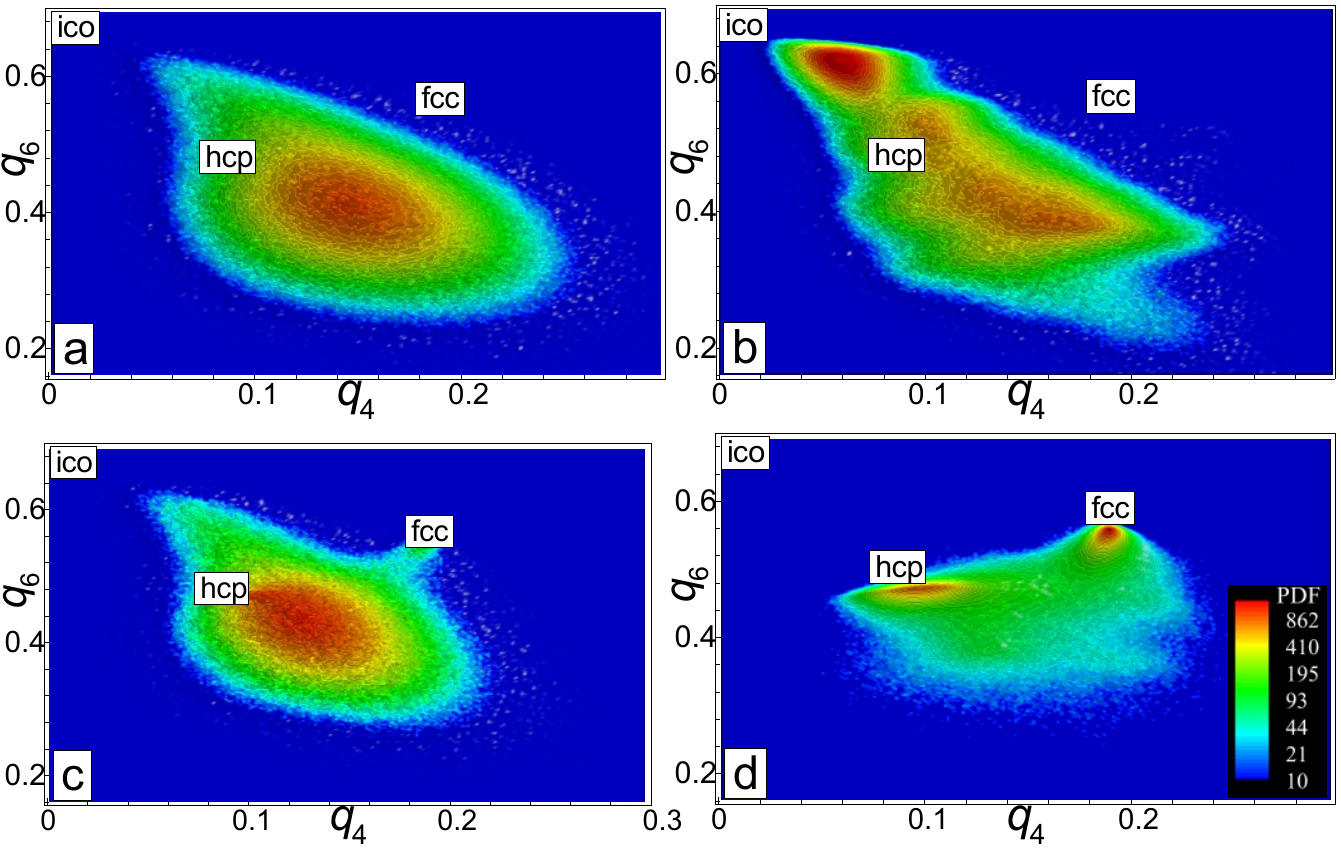}
  \\
  \caption{Illustration of  ``express'' method to estimate solid structure for given system parameters. The 2D probability distribution functions (PDF) plotted on the $(q_4,q_6)$-plane show dominant symmetry of local clusters even for badly relaxed structure whereupon the liquid-solid transition.  (a)-(b) show PDFs for RSS at ($\sigma=1.35$, $\rho=0.5$) for fluid at $T=0.103$ (a) and the ``poly-quasicrystalline'' solid (b) produced from the fluid after cooling to $T=0.1$. (c,d): the same for $\rho=0.45$  where the fluid after cooling from $T=0.12$ to $T=0.1$ transforms into polycrystal mixture of hcp and fcc-structures.}
  \label{fig4:q4q6}
\end{figure}

To define the local structural properties of the system, we use the bond order parameters method~\cite{Mitus1988PhysA,Steinhardt1981PRL,Steinhardt1983PRB,KlumovPU10,KlumovPRB11,Hirata2013Science} which is widely used to study different condensed matter systems. In this method, the rotational invariants of rank $l$ of both second $q_l({\bf r}_i)$ and third $w_l({\bf r}_i)$ order are calculated for each particle $i$ located at point ${\bf r}_i$ from the vectors (bonds) ${\bf r}_{ij}$ connecting its center with the centers of its $N_{\rm nn}({\bf r}_i)$ nearest neighboring particles:
\begin{gather}\label{wig}
q_l^2 ({\bf r}_i ) = \frac{{4\pi }}{{2l + 1}}\sum\limits_{m =  - l}^l {\left| {q_{lm} ({\bf r}_i )} \right|^2 }\,,
\\\notag
w_l ({\bf r}_i ) = \sum\limits_{\scriptstyle m_1 ,m_2 ,m_3  \hfill \atop
  \scriptstyle m_1  + m_2  + m_3  = 0 \hfill} {\left[ {\begin{array}{*{20}c}
   l & l & l \\
   {m_1 } & {m_2 } & {m_3 } \\
\end{array}} \right]} \;q_{lm_1 } ({\bf r}_i )q_{lm_2 } ({\bf r}_i )q_{lm_3 } ({\bf r}_i ),
\end{gather}
where $q_{lm}({\bf r}_i) = N_{\rm nn}({\bf r}_i)^{-1} \sum_{j=1}^{N_{\rm nn}({\bf r}_i)} Y_{lm}(\varphi({\bf r}_{ij}),\theta({\bf r}_{ij}))$, $Y_{lm}$ are the spherical harmonics and $(\varphi({\bf r}_{ij}),\theta({\bf r}_{ij}))$ are polar and azimuthal angles of the vectors ${\bf r}_{ij} = {\bf r}_i - {\bf r}_j$ connecting centers of particles $i$ and $j$. In Eq.(\ref{wig}) $[\ldots]$ are the Wigner 3$j$-symbols, and the summation  is performed over all the indexes $m_i =-l,...,l$, $i=1,2,3$.

The invariants $q_l$ and $w_l$ are uniquely determined for any crystalline structure and they are rotation invariant: this is the advantage of invariants. By varying number of nearest neighbors $N_{\rm nn}$ and rank $l$ of bond order parameter it is possible to identify {\it any} lattice type (including quasicrystalline particles and distorted  hcp/fcc/ico modifications, etc.) existing in the system. For more details of this procedure, see, e.g., Ref.~\cite{Ryltsev2013PRE}.

This method is the useful tool for express analysis of the structure. To demonstrate it we picture in Fig.~\ref{fig4:q4q6} the 2D probability distribution functions (PDF) plotted on the $(q_4,q_6)$-plane for RSS  at $\sigma=1.35$, $\rho=0.5$ and two temperatures which are slightly above Fig.~\ref{fig4:q4q6}(a) and slightly below Fig.~\ref{fig4:q4q6}(b)the liquid-QC transition temperature. The pictures show dominant symmetry of local clusters even for badly relaxed structure whereupon the liquid-solid transition.   It follows that the system local order after the transition is strongly icosahedral (10-fold tubes made of face-shared icosahedra, see Fig.~\ref{fig2:snapshots}(iii)). So the method is capable to distinct decagonal QC-like structures.  For comparison, we display pictures for $\rho=0.45$ (Figs.~\ref{fig4:q4q6}(b,c)),  where fluid after cooling transforms into polycrystal mixture of hcp and fcc-structures. Stable crystal structure at this density is fcc but, due to the fact that hcp structure has very close energy, the formation of such metastable mixtures is typical situation.

Of course the method can only determine local order; to study a global one we use diffraction analysis (see below).

 \begin{figure}[t]
  \centering
  \includegraphics[width=0.8\columnwidth]{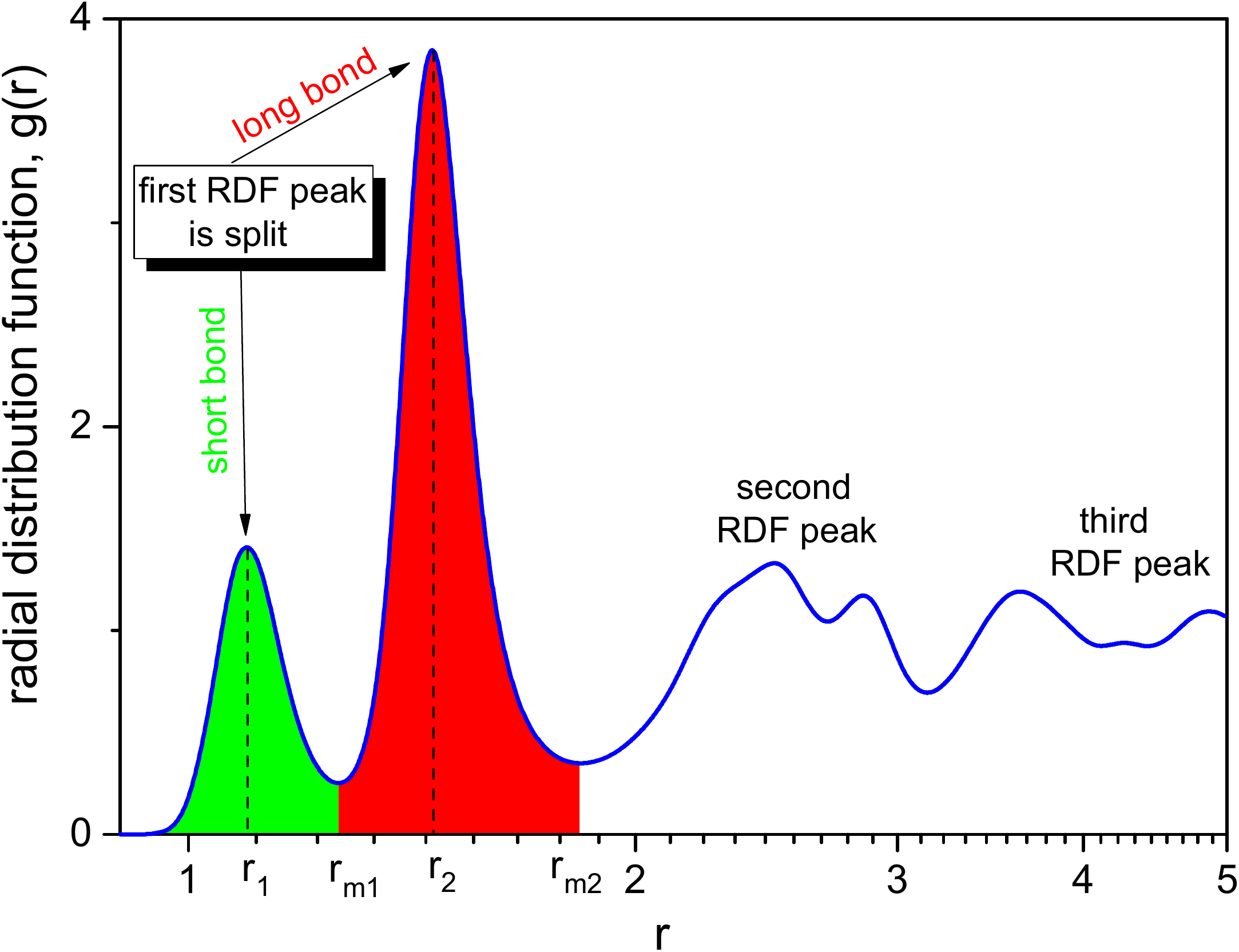}\\
  \caption{Demonstration of two-length-scale behaviour of the system under consideration. We see ``short'' and ``long'' interparticle bonds reflexed by splitting of the first radial distribution function peak. So we define two dimensionless parameters: the ratio between effective interparticle distances (bond lengthes), $\lambda$, and effective concentration of that bonds, $\phi$. The picture is obtained for RSS at $\sigma=1.35$, $\rho=0.51$, $T=0.1$ that corresponds to $\lambda=1.33$, $\phi=0.12$.}
  \label{fig3n:split}
\end{figure}

\subsection{Diffraction analysis}

To study global order of solid phases we use standard method of analyzing  reciprocal space structure which is equivalent to diffraction analysis. Within the frameworks of the method, we calculate the static structure factor
\begin{equation}\label{S(q)}
S({\bf k}) = \left\langle {\rho ({\bf k},t)\rho ( - {\bf k},t)} \right\rangle,
\end{equation}
where
$$
\rho ({\bf k},t) = \frac{1}{{\sqrt N }}\sum\limits_{i = 1}^N {e^{ - i{\bf kr}_i(t) } }
$$
is the spatial Fourier transform of particle density. Here ${\bf r}_i(t)$ is the instantaneous position of the $i$-th particle, ${\bf k}$ is a vector in the reciprocal space and $\left\langle {\cdots} \right\rangle$ denotes time averaging.

Projecting the particle coordinates onto a plane perpendicular to the high-symmetry directions (the symmetry axes) and calculating the structure factor (\ref{S(q)}) in the two-dimensional reciprocal space we obtain diffraction patterns such as those presented in Fig.~\ref{fig2:snapshots}(i,ii).

\section{Results and discussions}

\begin{figure}
  \centering
  \includegraphics[width=0.9\columnwidth]{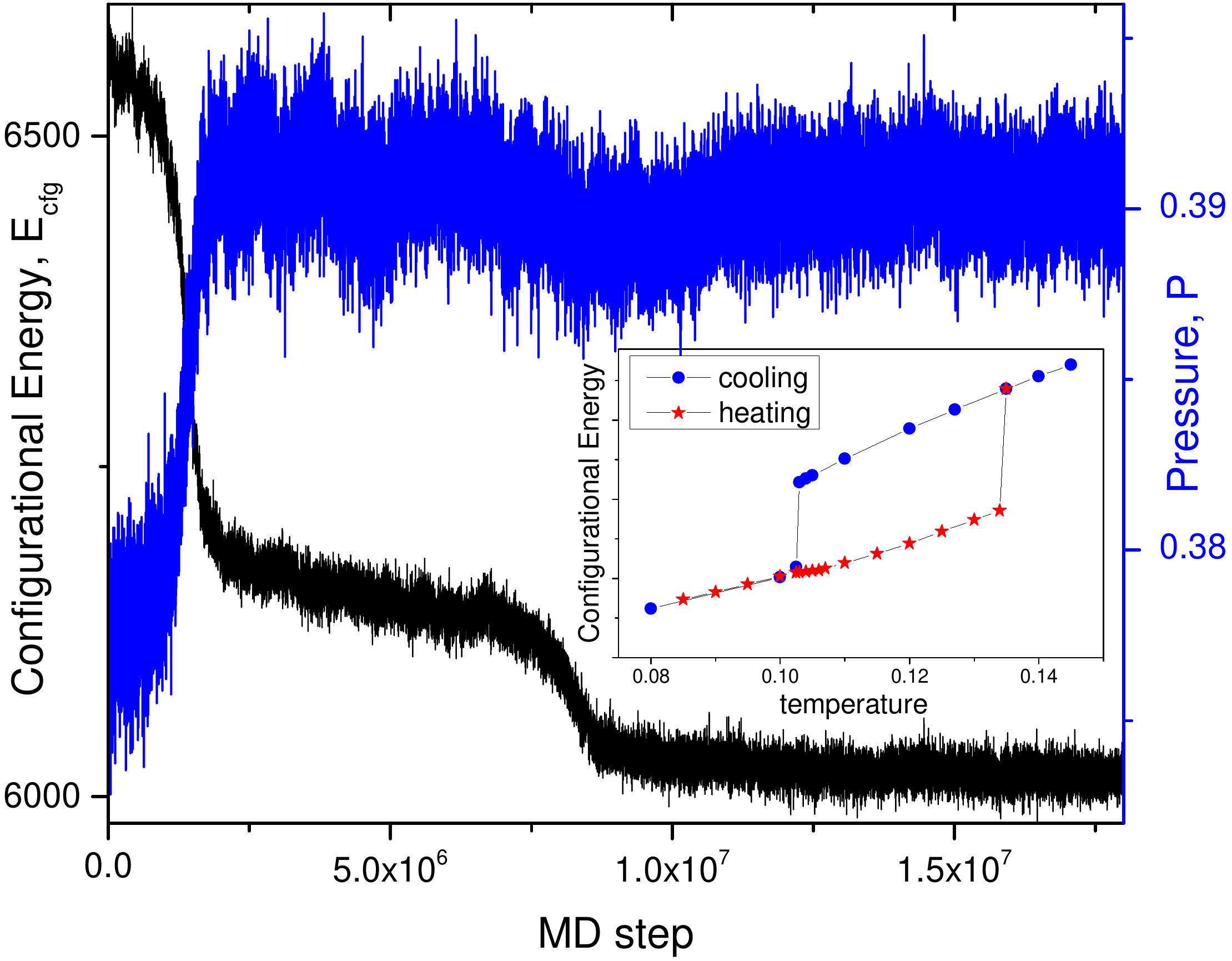}\\
 \caption{Evidence of first order scenario for liquid-QC transition in NVT ensemble simulations. Main frame: time evolution of configurational energy and pressure for system with RSS potential at $\sigma=1.37$, $\rho=0.474$, $T=0.1$ demonstrating sharp jumps at the transition. Insert: temperature dependencies of configurational energy of high temperature (fluid) phase at cooling and for low-temperature (QC) phase at heating demonstrating discontinuity and pronounced hysteresis. The dimensionless units are used for $E_{\rm cfg}$ and $P$ (see Sec.~\ref{sec:methods}).}
  \label{fig3:PT1}
\end{figure}

\subsection{Scale invariance and effective parameters}

\begin{figure*}[t]
  \centering
  \includegraphics[width=\textwidth]{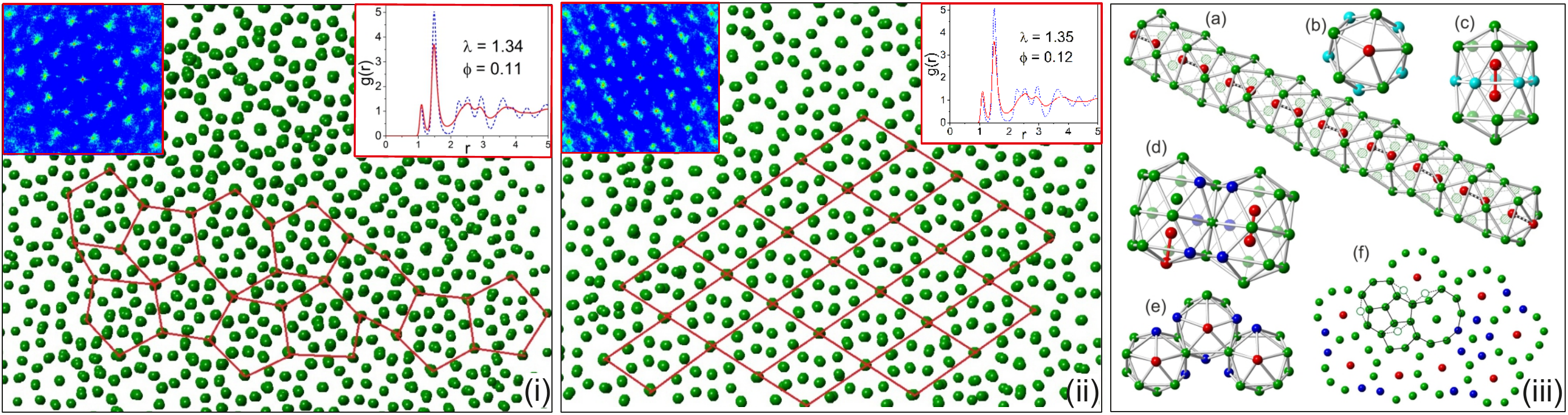}\\
  \caption{(i, ii) Typical snapshots of the RSS system demonstrating decagonal order. The pictures are shown in the plane perpendicular to the axes of decagonal tubes. Parameters values: (i) $\sigma=1.37$, $\rho=0.474$; (ii) $\sigma=1.38$, $\rho=0.47$. The values of $\lambda$ and $\phi$ are presented in the inserts. The lines connected the centers of decagons (or sites of alternative tiling elements) show different order at mesoscale: (i) demonstrates QC tiling whereas (ii) exhibits evident crystal symmetry. The inserts show diffraction patterns (left corners) and radial distribution functions (right corners) for fluid at $T=0.11$ (red solid) and solid at $T=0.08$ (blue dash); (iii) Structural elements of decagonal order:  (a) The structure of a decagonal tube made of face-shared icosahedra. Red particles located along the axis of the tube form short-bonded dimers. (b,c) Two face-shared icosahedra -- the building block of decagonal tubes. Particles belonging to common face are colored cyan. (d,e): The main junction mechanism of decagonal tubes -- edge sharing. The blue particles are also short-bonded. (f) The fragment of quasicrystal tiling in the plane perpendicular to the axes of tubes. Except decagons, tiling elements include pentagons, hexagons, rhombi and U-tiles (disturbed decagon). Solid and dashed lines mark the local rearrangements representing phason flips.}
  \label{fig2:snapshots}
\end{figure*}

Studying the RSS we observe a kind of scale invariance. Namely, the structure of low-temperature solid phases is essentially determined by two dimensionless parameters of the fluid that we cool down: the ratio between effective interparticle distances (bond lengthes), $\lambda$, and the fraction of short-bonded particles, $\phi$. Two-scale nature of the interaction potentials induces separation of the interparticle bonds on two sorts: ``short'' and ``long''.~\cite{Ryltsev2013PRL} That causes splitting of the first peak in the radial distribution function $g(r)$ (see Fig.~\ref{fig3n:split}). Thus $\lambda=r_2/r_1$, where $r_1$ and $r_2$ are the positions of the $g(r)$ subpeak maxima (Fig.~\ref{fig3n:split}).

Now we define, $r_{m1}$ and $r_{m2}$ --- locations of the first and the second $g(r)$ minima separating the subpeaks (Fig.~\ref{fig3n:split}).
The second effective parameter is the bond fraction $\phi=n_1/(n_1+n_2)$ , where $n_1  = 4\pi\rho \int_0^{r_{m1}} {r^2 g(r)dr}$ and $n _2  = 4\pi\rho \int_{r_{m1}}^{r_{m2}} {r^2 g(r)dr}$ are respectively the effective numbers of short- and long-bonded particles in the first coordination shell.

So further we use the effective parameters, $\lambda$ and $\phi$, as our frame of reference on top of $\rho$, $T$ and potential parameters. The scale invariance mentioned above survives while the fluid temperature is low enough so it has well defined local structure inherited from solid phase.~\cite{Ryltsev2013PRE,Ryltsev2014JCP} So $\lambda$ and $\phi$ should correspond to the temperature of a fluid  close to the transition to a solid phase.

First below we investigate the system with RSS potential and localize the range of the effective parameters $(\lambda,\phi)$ that favours QC order. Then we discuss QC order in systems with different potentials but the same values of effective parameters.

\subsection{Solid structures of RSS model}

For RSS system we mostly focus on the range of $\lambda\in(1.3,1.4)$ where non-trivial behavior has been already observed.~\cite{Fomin2008JCP,Gribova2009PRE,Ryltsev2013PRL} When $\phi<0.06$ and $\phi>0.3$ system crystalizes under cooling into  ``simple'' crystal phases like FCC, FCO, and SC structures. The range $\phi\in(0.15,0.3)$ corresponds to glassy state where crystallization is not observed during available simulation time scales due to frustrations caused by strong competition between different bond lengthes.~\cite{Ryltsev2013PRL} But within $\phi\in (0.06,0.15)$  situation is more complicated: system undergoes phase transition into QC with decagonal symmetry. Hereafter we use the term QC having in mind the system may also fall into crystalline approximant with local QC symmetry.~\cite{Goldman1993RevModPhys}  When $\phi\in(0.12-0.15)$ there is a phase ``intersection'': at intermediate cooling rates system falls into glassy state but at slower ones it undergoes the transition into QC.

Fig.~\ref{fig3:PT1} shows typical time dependencies of configurational energy $E_{\rm cfg}$ and pressure $P$ obtained in NVT ensemble simulations for RSS at the liquid-QC transition temperature. The transition is attended by sharp jumps of average values of $E_{\rm cfg}$ and $P$ which occur spontaneously at certain time. Pressure jump in NVT ensemble is  equivalent to density jump in NPT ensemble.  So there is release of heat and the density change at the transition. In the insert of Fig.~\ref{fig3:PT1}, temperature dependencies of $E_{\rm cfg}$ for both the high-temperature (liquid) and the low-temperature (QC) phases are respectively presented at cooling and heating. We see that system configurational energy undergos discontinuous change as the system is cooled below the certain transition temperature. Moreover, we see  pronounced hysteresis demonstrating strong supercooling (overheating) effect.  All of that gives a  reason to believe that the transition is of the first order. Note that both the $E_{\rm cfg}(t)$ and $P(t)$ dependencies reveal additional jumps taking places at longer time. That jumps are probaly related to structural relaxation of the low-temperature phase which contains a lot of defects whereupon the transition. Analysis of the configurations after the second jump just reveals better ordered structure with more pronounced decagonal order. So, most probably, there is no a second phase transition but only relaxation takes place.

Detailed analysis of the snapshots reveals that QC structure is ``polyquasicrystal'', i.e. it consists of well ordered QC grains. The snapshots in Fig.~\ref{fig2:snapshots}(i,ii) show the typical structures of the grains with pronounced decagonal (10-fold) symmetry. Decagons are in fact spatial ``tubes'' made of face-shared icosahedra (Fig.~\ref{fig2:snapshots}(iii)). The building block of such tubes is shown in Fig.~\ref{fig2:snapshots}(iii)b,c. As it is seen from Fig.~\ref{fig2:snapshots}(i,ii), icosahedral tubes tend to connect to each other and form QC-like mosaic in the plane perpendicular to tube axes. The main mechanism of this connection, edge sharing, is illustrated in Fig.~\ref{fig2:snapshots}(iii)d,e. The xyz-files of the snapshots presented in Fig.~\ref{fig2:snapshots} are available in Supplementary Information.

Note that decagonal structures presented in Fig.~\ref{fig2:snapshots} are essentially three-dimensional and cannot be treated as just stack of identical 2D layers. It can be seen from Fig.~\ref{fig2:snapshots}(iii) where the 3D-structure of QC building blocks is demonstrated. One can see that decagonal structures which appear as flat in Fig.~\ref{fig2:snapshots}(i,ii) are in fact build by particles belonging to three different planes (see differently colored particles in Fig.~\ref{fig2:snapshots}(iii)b). Moreover, considering  the atoms located along the axes of two adjacent decagonal tubes (Fig.~\ref{fig2:snapshots}(iii)d), we see that they form short-bonded dimers situated at different mutual positions. Because of 3D nature of decagonal structures observed, the most earlier obtained result and proposed methods concerning stability criteria of 2D QCs \cite{Barkan2011PRB,Engel2014PRL} hardly work in our case.

Despite of the same 10-fold symmetry at local scale, the structures illustrated in Fig.~\ref{fig2:snapshots}(i,ii) have different mesoscopic order corresponding to QC and approximant. Indeed, the lines connecting centers of decagons demonstrate random (quasi-periodic) tiling for the former case (Fig.~\ref{fig2:snapshots}(i)) but clear periodic structure for the later (Fig.~\ref{fig2:snapshots}(ii)). Diffraction patterns in the inserts confirm this conclusion. The most intensive diffraction peaks demonstrate 10-fold symmetry in both cases but additional peaks for the approximant pattern reveal clearly detectable crystalline symmetry. So RSS model demonstrates variety of phases with the same local decagonal symmetry but different mesoscopic order; the detailed investigation of their structure and thermodynamic stability is the matter of separate work.

We should also emphasize that decagonal structures presented in Fig.~\ref{fig2:snapshots}(i) is quasicrystalline in only two dimensions (the plain normal to axes of decagonal tubes) but periodic in the third dimension. This is the common feature of any layered (dodecagonal, octagonal) QCs.

 The snapshot corresponding to QC (Fig.~\ref{fig2:snapshots}(i)) shows tiling elements  alternative to decagons, such as rhombi, pentagons, hexagons and $U$-tiles \cite{Engel2007PRL}(disturbed decagons). The first three elements may serve as ``bricks'' filling the gaps between the decagons [rhombi also play this role in approximants, see Fig.~\ref{fig2:snapshots}(ii)]. $U$-tile as well as the combination of two pentagons and two hexagons represent examples of local arrangements which have energies close to those for decagon tile. Fig.~~\ref{fig2:snapshots}(iii)f shows typical structural rearrangements between such configurations (phason flips). The same tiles were observed for two-dementional decagonal QCs~\cite{Engel2007PRL,Kromer2012PRL}.

 Note that existence of phason strains impedes the study of global QC order due to distortion of diffraction patterns~\cite{Socolar1986PRB,Lubensky1986PRL,Goldman1993RevModPhys}.   Moreover, any QC-like configuration constrained by periodic boundary conditions is in fact a periodic approximant in the sense of global order. So we can discuss the difference between QC-like random tiling and crystalline approximants at only mesoscopic length-scales corresponding to MD box size. At that level, the difference between structures representing in Fig.~\ref{fig2:snapshots}(i,ii) is rather clear.

The presence of two characteristic interparticle distances plays the key role in the formation of the decagonal order. To show it we color by red and blue in Fig.~\ref{fig2:snapshots}(iii) the short-bonded nearest-neighbor particles. They are grouped in pairs along the axes of the icosahedral tubes (red) and at the centers of the pentagons belonging to adjacent icosahedra (blue). Without such short-bonded particles it is hardly possible to construct spatial decagonal clusters (like in Fig.~\ref{fig2:snapshots}). Such clusters are energetically disfavored in any system with simple one-scale potential like, e.g., Lennard-Jones one.

Surprisingly, despite of crucial role of short-bonded particles in QC formation, their fraction is relatively small. Indeed, the value $\phi\sim 0.1$ corresponds to about 1-2 short-bonded particles per first coordination shell. That means the decagonal QC formation can be favored by weak disturbance of an usual close packed system, see discussion in Sec.~\ref{sec:experiment}.

\subsection{Criterion for decagonal order formation and validation}

So there are two conditions necessary for formation of 3D QC structure: 1) optimal ratio $\lambda$ between the lengthes of short and long nearest-neighbor bonds to minimise the icosahedral distortion and 2) optimal  fraction of short-bonded particles $\phi$. For decagonal QC-like order, we estimate the optimal parameters of $\lambda\simeq(1.35-1.4)$ and $\phi\simeq (0.06-0.15)$. These conditions can be  satisfied in relatively small manifold of system parameters.

To validate the criterion proposed, we perform simulations with two alternative two-scale potentials (see Fig.~\ref{fig1:RDF}a). Adjusting system parameters to obtain appropriate values of $(\lambda, \phi)$ in liquid state (Fig.~\ref{fig1:RDF}b), we cool the systems and observe the self-assembly of similar decagonal solid phases (see snapshots in Fig.~\ref{fig1:RDF}b). Note that these phases are observed at temperature-density ranges which are essentially different from those for RSS ($\rho=0.74$, $T=0.085$ for YK and $\rho=0.62$, $T=0.15$ for OPP). That suggests proposed criterion of decagonal structure formation is general and does not depend on any peculiarities of the system except the existence of two length-scales of the interaction. Of course, some more subtle features of decagonal phases such as the regions of QC stability or the structure of competing approximants may depend on particular system properties.

Thus we have shown that structure of solid QC phase can be effectively predicted by two simple structural parameters of fluid phase. Their estimation is much easier task than solid structure determination that allows quickly localizing the area of system parameters where one may expect decagonal order formation. Indeed, the value of ratio between short and long bond lengthes $\lambda$ is easily estimated from the pair potential. So, for a system with attractive wells, $\lambda$ is equal to the ratio of the locations of potential minima (see Fig.~\ref{fig1:RDF} and Ref.~\cite{Englel2015Nature}); for purely repulsive potentials $\lambda$ can be estimated through the width of the repulsive shoulder (for example, $\lambda\sim\sigma$ for RSS). The value of short bond concentration $\phi$ mostly depends on system density and so it can be evaluated during quasi-equilibrium MD run in NPT ensemble with varying barostat pressure.


\subsection{Relation to experiment\label{sec:experiment}}
There is a discrepancy between soft matter system with quasiperiodic order and metallic quasicrystals. Metallic quasicrystals exhibit predominantly decagonal and icosahedral long-range order, while soft matter systems so far appeared to prefer dodecagonal order. Our results demonstrate there is no fundamental reason why decagonal QCs should not also be possible in soft matter. We show that decagonal order takes place if $\lambda \sim 1.35$ and $\phi \sim 0.1$ that means a small fraction of short-bonded particles (about 1-2  particles per first coordination shell) with relatively small bond-length difference. The known case of computing self-assembly of 3D dodecagonal QCs \cite{Dzugutov1993PRL} suggests that 12-fold symmetry is favored at essentially different  $\lambda -\phi$ domain around  $\lambda \sim 1.7$, $\phi \sim 0.4$.

So we suggest that soft matter QCs with decagonal symmetry can be experimentally observed for the same systems demonstrating the dodecagonal order for an appropriate choice of parameters controlling $\lambda$ and $\phi$.

The promising soft matter systems to perform such experiments are polymer micelles for which 12-fold QC order was recently observed.~\cite{Fischer2011PNAS} Such micelles exhibit a soft interaction potential inherently characterized by two length scales, i.e., the dimensions of the micellar core and the micellar shell.~\cite{Fischer2011PNAS} So it is possible to tune effective parameters $\lambda$ and $\phi$ by varying, for example, the core/shall ratio and density of the micelles.

\section{Conclusions}

To conclude, we first show decagonal QC order in simple one component 3D system and demonstrate that even purely repulsive soft-matter-like interaction can produce such type of QCs. The underlined mechanism of decagonal order formation is the stabilization of the tube-like clusters made of face-shared icosahedra due to the existence of two effective interparticle distances. We propose the criterion which allows predicting decagonal QC formation by means of two simple structural parameters of fluid phase derived from its radial distribution function.  We demonstrate validity of the criterion on three different two-scale systems.  Our work is not the last word in the problem but has raised a lot of open issues whose solution will help understanding nature of QC states observed experimentally in molecular liquids, soft matter systems and metallic alloys.

\section{Acknowledgments}
We have come to the idea of universality doing investigation of CuZr glass forming ability (within Russian Scientific Foundation grant No 14-13-00676) where pair potentials are multiscale as in this paper. Within Cu-Zr investigation we also developed methods of orientational order analysis used here. So we thank grant RSF No 14-13-00676.  Molecular dynamic simulations on supercomputers and most part of the data processing have been performed using support of RSF grant No. 14-12-01185. We are grateful to Russian Academy of Sciences for the access to JSCC and ``Uran'' clusters. We thank S.~Buldyrev and M.~Engel for helpful discussions.




\bibliography{our_bib_rss} 
\bibliographystyle{rsc} 

\end{document}